\documentclass[preprint,12pt]{elsarticle}
\usepackage{algorithm}
\usepackage{hyperref}
\usepackage{algpseudocode}
\usepackage{amsmath}
\usepackage{mathtools}
\usepackage{amsmath}
\usepackage{amsthm}
\usepackage{multirow}
\usepackage{longtable}
\usepackage{array}
\usepackage{xcolor}
\usepackage{tabularx}
\newcommand{\colorB}{\color{blue}}
\newcommand{\colorR}{\color{red}}
\newcommand{\colorBB}{\color{blue}}
\newcommand{\colorRR}{\color{red}}

\def\one{\leavevmode\hbox{\small1\normalsize\kern-.33em1}}

\newcommand{\num}{d}
\DeclareMathOperator\sgn{sgn}

\usepackage{amssymb}

\newcounter{bla}

\journal{Computer Physics Communications}

\begin{document}
\begin{frontmatter}

\title{Classical bounds on two-outcome bipartite Bell expressions and linear prepare-and-measure witnesses: Efficient computation in parallel environments such as graphics processing units}

\author[a]{István Márton\corref{author}}
\author[a]{Erika Bene}
\author[b]{Péter Diviánszky}
\author[a,c]{G\'abor Dr\'otos}

\cortext[author] {Corresponding author.\\\textit{E-mail address:} marton.istvan@atomki.hu}
\address[a]{HUN-REN Institute for Nuclear Research, P.O. Box 51, Debrecen, H-4001, Hungary}
\address[b]{Faulhorn Labs, Budafoki út 91-93, Budapest, H-4117, Hungary}
\address[c]{Instituto de F\'{i}sica Interdisciplinar y Sistemas Complejos (IFISC), CSIC-UIB, Campus UIB, Carretera de Valldemossa, km 7,5, E-07122 Palma de Mallorca, Spain}

\begin{abstract}

The presented program aims at speeding up the brute force computation of the so-called $L_\num$ norm of a matrix $M$ using graphics processing units (GPUs). Alternatives for CPUs have also been implemented, and the algorithm is applicable to any parallel environment. The $n\times m$ matrix $M$ has real elements which may represent coefficients of a bipartite Bell expression or those of a linear prepare-and-measure (PM) witness. In this interpretation, the $L_1$ norm is the local bound of the given correlation-type Bell expression, and the $L_\num$ norm for $d\ge 2$ is the classical $d$-dimensional bound of the given PM witness, which is associated with the communication of $d$-level classical messages in the PM scenario. The program is also capable of calculating the local bound of Bell expressions including marginals. In all scenarios, the output is assumed to be binary.

The code for GPUs is written in CUDA C and can utilize one NVIDIA GPU in a computer. To illustrate the performance of our implementation, we refer to Brierley et al.~\cite{BrierleyArxiv2017} who needed approximately three weeks to compute the local bound on a Bell expression defined by a $42\times 42$ matrix on a standard desktop using a single CPU core. In contrast, our efficient implementation of the brute force algorithm allows us to reduce this to three minutes using a single NVIDIA RTX 6000 Ada graphics card on a workstation. For CPUs, the algorithm was implemented with OpenMP and MPI according to the shared and distributed memory models, respectively, and achieves a comparable speedup at a number of CPU cores around 100.
\\

\noindent\textbf{PROGRAM SUMMARY}
\begin{small}
\noindent
{\em Program Title:} L\textunderscore CUDA.cu, L\textunderscore MPI.c, L\textunderscore OpenMP.c \\
{\em CPC Library link to program files:} (to be added by Technical Editor) \\
{\em Developer's repository link: } \url{https://github.com/istvanmarton/L-norms_BruteForce} \\ 
{\em Code Ocean capsule:} (to be added by Technical Editor)\\
{\em Licensing provisions:} GPLv3\\
{\em Programming language:} C, CUDA, OpenMP, MPI \\
{\em Nature of problem(approx. 50-250 words):}\\
  The computational demand of determining the $L_\num$ norm of a matrix of real coefficients is high; exact $L_\num$ norms have been computed so far for relatively small matrices only. Besides that any exact algorithm appears to scale exponentially with the number of rows (or the minimum of rows and columns, for $\num = 1$), the naive approach to brute-force compute an $L_\num$ norm features a contribution that scales linearly with the product of the number of rows and columns, because every matrix element needs to be accessed repeatedly. Improving efficiency is thus both desirable and possibly feasible.
  \\
{\em Solution method(approx. 50-250 words):}\\
  We attenuate the mentioned secondary contribution by accessing elements only from a single row of the matrix in each step, achieving a linear scaling with the number of columns in terms of memory access. Even though we perform a search for the relevant row scaling linearly with the number of rows in terms of the number of operations, this precedes the access to the matrix elements, so that multiplication does not occur for time complexity. Besides identifying further important mathematical shortcuts, our algorithm is very well suited to parallelization, which we take advantage of. In particular, we provide an implementation for graphics processing units besides universal processors.
  \\
{\em Additional comments including restrictions and unusual features (approx. 50-250 words):}\\
  The program can utilize a single GPU on a desktop, and the entries of the input matrix $M$ must be provided as integers (i.e., conversion from floating-point entries is necessary).
  \\
\end{small}
\end{abstract}
\end{frontmatter}

\section{Introduction}
In many applications of quantum information theory, such as quantum communication tasks and quantum cryptographic protocols~\cite{Nielsen2010}, black box devices are arranged in general communication networks that can exchange and process information~\cite{Scarani2019}. In these frameworks, quantum dimensionality is an important resource which allows quantum protocols to largely outperform classical ones of fixed dimensionality~\cite{BrunnerPRL2008,GallegoPRL2010}.  

One of the main questions in this field is how to assess the dimension of classical or quantum states from the observed statistics alone, without making any assumptions about the devices used in the protocol (see references~\cite{BrunnerPRL2008, GallegoPRL2010, PalPRA2008, WehnerPRA2008, WolfPRL2009} for early works). This is the so-called device-independent approach to quantum information; along with its semi-device-independent relaxation which allows for some minimal assumptions (often called black-box framework together). It has applications in quantum cryptography~\cite{Scarani2009, Woodhead2015}, quantum randomness generation~\cite{AcinNature2016, LiPRA2012}, quantum communication complexity~\cite{AmbainisJACM2002} and beyond~\cite{ScaraniActPhysSlov2012}. 

Gallego et~al.~\cite{GallegoPRL2010} developed a general device-independent framework to test the classical and quantum dimensionality. Using geometric considerations, tight classical dimension witnesses were constructed and extended to witnesses of quantum dimension. Later, this framework was generalized to arbitrary quantum networks~\cite{BowlesPRA2015}. These dimension witnesses give lower bounds on the classical dimension and the Hilbert space dimension required to reproduce the observed measurement statistics, respectively. These tests have also been implemented in laboratory experiments. Hendrych et~al.~\cite{HendrychNatPhys2012} carried out experiments with photon pairs entangled in both polarization and orbital angular momentum, whereas Ahrens et~al.~\cite{AhrensNatPhys2012} encoded information in polarization and spatial modes to demonstrate the power of dimension witnesses.

Beyond dimensionality, however, dimension witnesses can also serve as criteria to distinguish between classical and quantum resources in black-box tests. In the prepare-and-measure scenario, with a large number of preparation and measurement setting input values and assuming a message of dimension two (bit case), Diviánszky et~al.~\cite{DivianszkySciRep2023} demonstrated quantumness by bounding the classical dimension witness using large-scale numerical tools. Based on this one-bit bound, new constants were defined which were related to the Grothendieck constant of order 3~\cite{Grothendieck1953, AcinPRA2006, DivianszkyPRA2017} and were associated with the white noise tolerance of the prepared qubits and the critical detection efficiency of the measurements. Such tasks are, however, computationally costly.

In this paper, we present efficient brute force computations of both the local bound of Bell expressions and the one-dit classical bound of linear witnesses in the prepare-and-measure scenario using parallel environments such as graphics processing units (GPUs). The local (classical) bound of a generic correlation-type Bell expression, irrespective of the dimensionality, is given by~\cite{BrierleyArxiv2017,DivianszkyPRA2017, BrunnerRevModPhys2014}
\begin{equation}
\label{LM2}
L_1(M):=\max_{\{a_x=\pm1\}}\sum_{y=1}^m\left|\sum_{x=1}^n M_{xy}a_x\right| = \max_{\{a_x=\pm1\}} \lVert \mathbf{a}M \rVert_1 =: \max_{\{a_x=\pm1\}} L^*_1(M;\mathbf{a}) ,
\end{equation}
where $M$ is an $n\times m$ matrix of real coefficients; this is how the $L_1$ norm of $M$ is defined. Here, $\lVert \mathbf{v} \rVert_1$ is the Manhattan norm for an arbitrary real vector $\mathbf{v}$, i.e., $\lVert \mathbf{v} \rVert_1 = \sum_i \left| v_i \right|$. For later use, we have introduced in the last step what we call an $L^*_1$ Bell value (i.e., a possible value of the corresponding Bell expression), and which is just the Manhattan norm of the vector $\mathbf{m} := \mathbf{a}M$ for a given matrix $M$ and vector $\mathbf{a}$. The vector $\mathbf{a}$ is called a strategy vector according to a game theoretic approach to Bell expressions~\cite{PalazuelosJofMathPhys2016}. Beyond Bell-like scenarios, the $L_1$ norm of a matrix also has significance in the theory of Grothendieck's constant~\cite{Raghavendra2009}, in the theory of graphs~\cite{FriezeCombinatorica1999, BorgsAdvinMath2008} or in communication complexity~\cite{LinialCombinatorica2007}.

In a generic Bell expression Alices's (Bob's) marginal coefficients are defined by the first row (column) of the Bell matrix \cite{GohPRA2018, SliwaPhysLettA2003, AraujoArxiv2024}. It is easy to show that the local bound can be expressed in this case as
\begin{equation}
\label{LMarg}
L_{marg}(M):=\max_{\left\{\substack{a_1 = +1 \\ a_x = \pm 1,\text{ } x \ge 2} \right\}} \left(\sum_{x=1}^n M_{x1}a_x + \sum_{y=2}^m\left|\sum_{x=1}^n M_{xy}a_x\right| \right).
\end{equation}

In the case of the prepare-and-measure scenario, Alice prepares a quantum state $\rho_x$ upon receiving an input $x\in \{1,\ldots,n\}$ and sends it to Bob. Bob measures the prepared state $\rho_x$ in a measurement setting chosen as a function of an input $y \in \{1,\ldots,m\}$, and obtains an outcome $b\in\{-1,1\}$ with the conditional probability $P(b\mid x,y)$. With the coefficients $M_{xy}$ of an arbitrary real $n \times m$ dimensional witness matrix, we can define the quantity called witness as
\begin{equation}
W=\sum_{x=1}^{n}\sum_{y=1}^{m} M_{xy}E_{xy} ,
\label{dimwit}
\end{equation}
where $E_{xy} = P(b=+1\mid xy)-P(b=-1\mid xy)$ are expectation values.

If Alice can only prepare classical $d$-dimensional systems (dits, with $d\geq 2$), we denote the upper bound of Eq.~(\ref{dimwit}) by $L_\num(M)$ and call it the $L_\num$ norm of $M$:
\begin{equation}
L_\num(M):=\max\sum_{x=1}^{n}\sum_{y=1}^{m}M_{xy}E_{xy},
\label{L2M}
\end{equation}
where we maximize over all classical expectation values that can be attained with $d$-dimensional systems. Note that we cannot derive Eq.~(\ref{LM2}) by choosing $\num = 1$, namely, Eq.~(\ref{LM2}) is not a special case of Eq.~(\ref{L2M}).

\begin{figure}[h]
\centering\includegraphics[width=14.0cm]{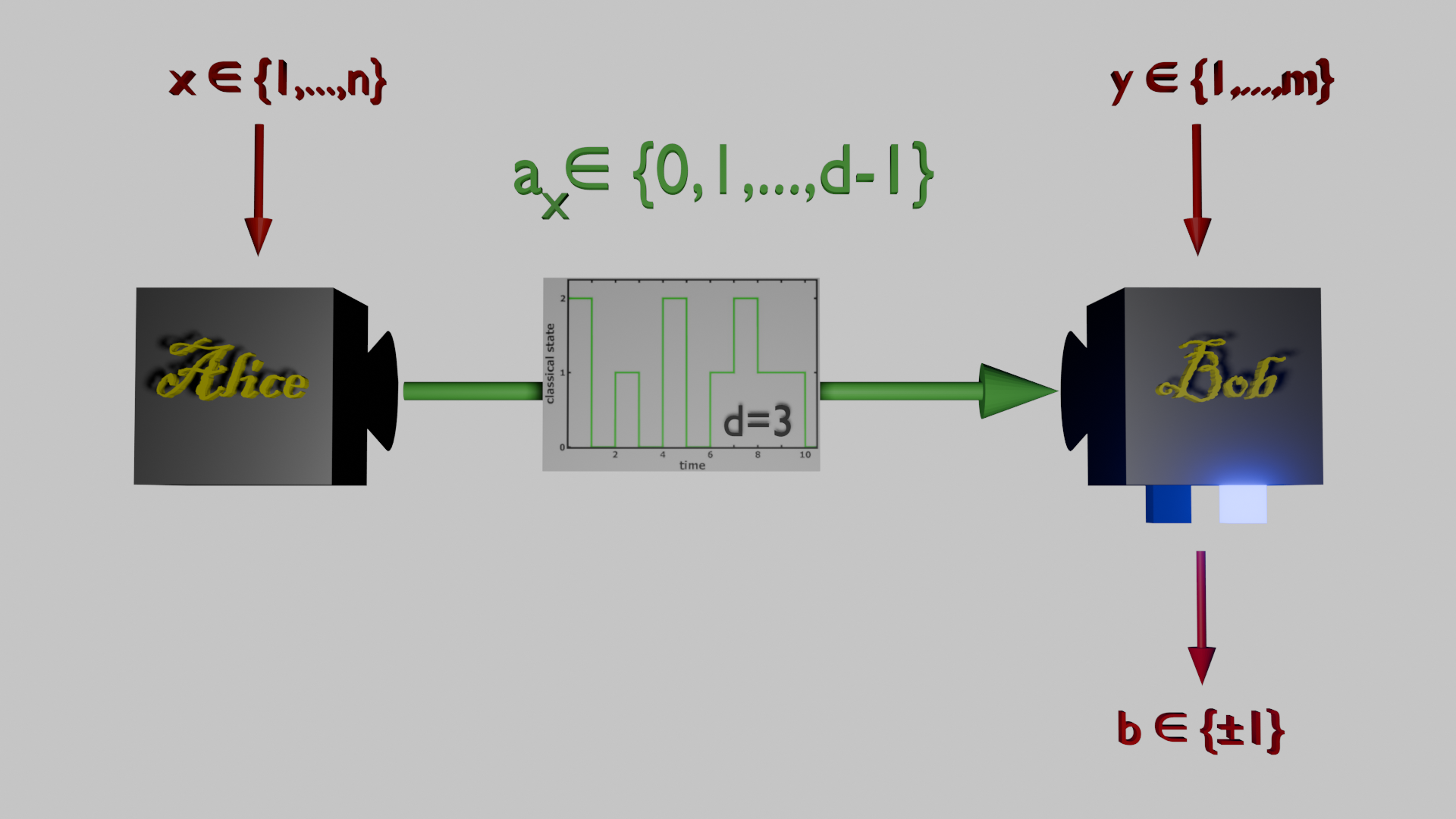}
\caption{The prepare-and-measure setup using a classical dit of communication. The inset in the middle illustrates an implementation of the $d=3$ case relying on a classical signal taking values $a\in\{0,1,2\}$ and varying in time to distinguish between subsequent rounds of communication.} \label{AliceBobfig}
\end{figure}
In fact, the maximum can be achieved by deterministic protocols when Alice, depending on $x$, sends a classical message $a_x \in \{ 0, 1, \cdots, \num-1\}$ to Bob, whose binary output $b \in \{-1,1\}$ is a deterministic function of $a_x$ and $y$ (see Fig.~\ref{AliceBobfig}). Let us denote Bob's output by $b_y^a$ if $a_x = a$, i.e., if Alice's message $a_x$ takes some given value $a$; then $E_{xy} = b_y^a$ if $a_x = a$, $a \in \{ 0, 1, \cdots, \num - 1\}$. According to these considerations, Eq.~(\ref{L2M}) can be written as
\begin{align}
&L_\num(M)=\nonumber\\
&=\max_{\substack{\{a_x=0,+1,\cdots, \num-1\} \\ \{b_y^a = \pm1\}}} \left(\sum_{x: a_x=0}\sum_{y=1}^{m} M_{xy}b^0_y+\sum_{x: a_x=1}\sum_{y=1}^{m} M_{xy}b^1_y + \dots + \sum_{x: a_x=\num-1}\sum_{y=1}^{m} M_{xy}b^{\num-1}_y\right).
\label{L2Mv2}
\end{align}

Note that the $b_y^a$ variables can be eliminated, and we can eventually write
\begin{equation}
\begin{split}
    L_\num(M)
    =\max_{\{a_x=0,+1,\cdots, \num-1\}} \sum_{a=0}^{\num-1} \lVert \mathbf{m}_a \rVert_1 =: \max_{\{a_x=0,+1,\cdots, \num-1\}} L^*_\num(M;\mathbf{a}) ,
    \label{L_o_norm}
\end{split}
\end{equation}
where we define the components of vectors $\mathbf{m}_a$ for each $a\in\{0, 1, \cdots, \num-1\}$ by
\begin{equation}
	\left(\mathbf{m}_a\right)_y = \sum_{x: a_x=a} M_{xy} .
	\label{eq:ma}
\end{equation}
We also introduce $L^*_\num$ witness values in the last step of Eq.~(\ref{L_o_norm}). These are the sums of the Manhattan norms of all different vectors $\mathbf{m}_a$, $a\in\{0, 1, \cdots, \num-1\}$.

$L_\num$, for any integer $\num \geq 1$, has several interesting properties~\cite{DivianszkySciRep2023}. Notably, it can be proved that $L_\num$ is a matrix norm. In~\cite{DivianszkySciRep2023} two efficient numerical algorithms were used to compute $L_2(M)$ for large matrices $M$: a heuristic see-saw-type algorithm usually giving a tight lower bound on $L_2(M)$ was verified by a branch-and-bound-type algorithm giving the exact value of $L_2(M)$. However, for certain matrices the running time of the branch-and-bound method is too large. On the other hand, the computational time of the brute force method is more predictable for $L_d$.

In the next section, we will describe the techniques to speed up the computation of the $L_\num$ norms of a given matrix. We have used both mathematical/algorithmic considerations and parallel computing to achieve this goal. The codes are written in the C programming language with either one of the following three parallel computation environments: OpenMP, MPI, or CUDA. OpenMP and MPI are designed to use shared memory and distributed memory parallelism, respectively \cite{Pacheco2011}. CUDA is designed for the programming of Graphics Processing Units (GPUs) \cite{HanCUDA2019}. GPUs are famous for having generally several thousands of execution units, making them ideal for computational tasks in a massively parallel environment.

\section{Description of the method}

\subsection{$L_1$ norm}\label{L1_norm_subsection}
Consider a matrix $M$ with $n$ rows and $m$ columns. Consider a strategy vector $\mathbf{a}$ of dimension $n$, where the vector entries can be $+1$ or $-1$ resulting in $2^n$ possible options for the vector $\mathbf{a}$. The $L_1$ norm of $M$ was defined in Eq.~(\ref{LM2}) through all those strategy vectors as the optimal value among the Manhattan norms of the vectors $\mathbf{m} = \mathbf{a}M$. Computing the $L_1$ norm naively would thus mean $2^n$ matrix multiplications with a vector. We have identified four ways to reduce the number of operations to calculate the $L_1$ value. Two of them correspond to a prior transformation of the matrix, while the other two concern the actual algorithm representing the brute-force computation. We will discuss them all below, before describing parallelization.

\subsubsection{Preprocessing the matrix}
The first simplification is a possible reduction of the number of rows or columns of the matrix such that the $L_1$ norm is not affected. We have found two cases when such a reduction can actually be performed:
\begin{itemize}
  \item Trivially, when all entries of a given row or column are zero; such a row or column can be removed.
  \item When a row or column of the matrix is a constant multiple of another row or column. In this case, the two rows or columns can be replaced by a single one by adding or subtracting the two, depending on the sign of the constant multiplier.
\end{itemize}

These opportunities for reduction are proved in \ref{AppB} and are illustrated on the example of the following matrix:
\begin{equation}
M = 
\begin{pmatrix*}[r]
0 & 0 & 0 & 0\\
4 & -7 & 2 & -1\\
8 & -14 & 4 & -2\\
1 & -3 & 4 & 4
\end{pmatrix*}
\longrightarrow
\begin{pmatrix*}[r]
12 & -21 & 6 & -3 \\
1 & -3 & 4 & 4
\end{pmatrix*}
.
\end{equation}
We can see that the first row does not influence the $L_1$ norm of the matrix and that the third row can be added to the second.

The other prior transformation utilizes the special property of the $L_1$ norm that the transposed matrix has the same $L_1$ norm as the original one. This feature can be taken advantage of because the computational costs scale more expensively with the number of rows. Accordingly, the program transposes the matrix when the number of rows exceeds the number of columns.

\subsubsection{Algorithmic speedups}
An immediate algorithmic simplification arises from the fact that the $L^*_1$ Bell values are the same for a given strategy $\mathbf{a}$ and for the opposite strategy vector, $-\mathbf{a}$. Then all possible $L^*_1$ Bell values will remain covered if we fix one entry of the strategy vector. The number of strategy vectors to be considered for calculating all possible $L^*_1$ Bell values thereby decreases to $2^{n-1}$. (Note that this already implies that the optimal strategy vector is not unique as the optimal $L^*_1$ Bell value can be attained by two different strategy vectors at least; but even more strategy vectors may be optimal in other cases.)

Perhaps the most important speedup, however, is based on the idea that we do not need to perform a full vector-matrix multiplication to obtain each $L^*_1$ Bell value apart from the first one. Instead, we can use a series of strategy vectors where the neighboring vectors have a Hamming distance of one. After calculating the $L^*_1$ Bell value for the first strategy vector, each subsequent $L^*_1$ Bell value can be calculated from the previous one by accessing the specific row of the matrix where the value of the strategy vector is modified, and all possible $L^*_1$ Bell values can thus be obtained from the first one step by step. Note, however, that one needs to store each individual component of the vector $\mathbf{m} = \mathbf{a}M$ between consecutive steps (i.e., between subsequent strategy vectors $\mathbf{a}$).

To better understand, let us consider the following matrix:
\begin{equation}
M = 
\begin{pmatrix*}[r]
4 & -7 & -2\\
-5 & 2 & 3 \\
9 & -1 & 4
\end{pmatrix*}
.
\end{equation}
The $L^*_1$ Bell values corresponding to the strategy vectors displayed on the right below should be calculated as follows. The value in the first line is calculated by performing the vector-matrix multiplication $\mathbf{m}^{(1)} = \mathbf{a}^{(1)}M$, then taking the Manhattan norm of $\mathbf{m}^{(1)}$. Each subsequent line modifies the previous $L^*_1$ Bell value by updating the components of the vector $\mathbf{m}$ according to the change in the strategy vector $\mathbf{a}$.
 \begin{align} 
    \label{L1_calculate_simplified_method}
    &L^*_1(M,\mathbf{a}^{(1)}) = \left|-8\right| + \left|6\right| + \left|-5\right| = 19 & (-1, -1, -1)\\
    \label{L1_calculate_simplified_method_second}
    &L^*_1(M,\mathbf{a}^{(2)}) = \left|-8 + 2 \times 9\right| + \left|6 + 2 \times \left(-1 \right)\right| + \left|-5 + 2 \times 4 \right| = 17 & (-1, -1, +1)\\ 
    \label{L1_calculate_simplified_method_third}
    &L^*_1(M,\mathbf{a}^{(3)}) = \left|10 + 2 \times \left(-5 \right) \right| + \left|4 + 2 \times 2 \right| + \left|3 + 2 \times 3\right| = 17 & (-1, +1, +1)
 \end{align}
 \vspace{-0.6cm}
    \begin{equation}
        \vdots \nonumber
    \end{equation}

To describe this method we need to consider a series of vectors having a Hamming distance of one like the column vectors in Table~\ref{Hamming_table}. A simple choice is the so-called binary reflected Gray code (BRGC) \cite{BitnerCACM1976, ConderDiscMath1999}; see Eqs.~(\ref{N_ary_G1})-(\ref{N_ary_G3}) for a definition in case $d=2$, and refer to Table~\ref{Hamming_nplus1} for a recursive and more pictorial formulation. Note that the BRGC entries $G$ take a value of $0$ or $1$ according to the definition; however, conversion to entries $a = 1$ or $a = -1$ is straightforward through $a=2G-1$, which results in a one-to-one correspondence between the strategy vectors and the columns of the BRGC.

To successfully implement this method (especially in a parallel environment) we need to be able to calculate the entries $G$ of the BRGC as a function of the position $(i,j)$, where $i$ is the index of the vector component (digit) of the $j$th vector (word) of the BRGC. By inspecting such a series of vectors in the columns of Table~\ref{Hamming_table}, we can notice that the BRGC can be described with exponents of two determined by $i$. In particular, all rows start with $2^{i}$ zeros. When $i$ is maximal, the zeros are followed by $2^{i}$ ones. When $i$ is not maximal, the zeros are followed by $2^{i+1}$ consecutive ones and then again by zeros until the last sequence of $2^{i}$ zeros. This pattern is depicted in Table~\ref{Hamming_table_pattern}. We can easily observe that if we increase $j$ by $2^{i}$ and then divide it by $2^{i+1}$, whether the integer part will result in an even or an odd number will depend on the value of the vector at the given row. In fact, if we take the remainder of this number with respect to division by 2, we just obtain the entry of the given, $j$th, vector in the given, $i$th, row (the $i$th digit of the $j$th word). That is, the entries $G$ of the BRGC can be calculated as follows:
\begin{equation}
    G_{i,j} = \left\lfloor\frac{j + 2^{i}}{2^{i + 1}}\right\rfloor \mod 2.
    \label{Hamming_equation}
\end{equation}
See \ref{Appendix_1} for a proof.

\begin{table} []
\setlength{\tabcolsep}{4.5pt}
\renewcommand{\arraystretch}{1.0}
\begin{tabular}{rccccccccccccccccc}
\multicolumn{16}{c}{$j$th word} \\

\vline & 0 & 1 & 2 & 3 & 4 & 5 & 6 & 7 & 8 & 9 & 10 & 11 & 12 & 13 & 14 & 15 \\
\hline
\multirow{4}{1.2em}{\rotatebox[origin=r]{90}{$i$th digit}} 3 \vline & \colorB 0 & \colorB 0 & \colorB 0 & \colorB 0 & \colorB 0 & \colorB 0 & \colorB 0 & \colorB 0 & \colorR 1 & \colorR 1 & \colorR 1 & \colorR 1 & \colorR 1 & \colorR 1 & \colorR 1 & \colorR 1 \\
 2 \vline & \colorB 0 & \colorB 0 & \colorB 0 & \colorB 0 & \colorR 1 & \colorR 1 & \colorR 1 & \colorR 1 & \colorR 1 & \colorR 1 & \colorR 1 & \colorR 1 & \colorB 0 & \colorB 0 & \colorB 0 & \colorB 0 \\
 1 \vline & \colorB 0 & \colorB 0 & \colorR 1 & \colorR 1 & \colorR 1 & \colorR 1 & \colorB 0 & \colorB 0 & \colorB 0 & \colorB 0 & \colorR 1 & \colorR 1 & \colorR 1 & \colorR 1 & \colorB 0 & \colorB 0 \\
 0 \vline & \colorB 0 & \colorR 1 & \colorR 1 & \colorB 0 & \colorB 0 & \colorR 1 & \colorR 1 & \colorB 0 & \colorB 0 & \colorR 1 & \colorR 1 & \colorB 0 & \colorB 0 & \colorR 1 & \colorR 1 & \colorB 0 \\
\hline
& 3$^-$ & 0$^+$ & 1$^+$ & 0$^-$ & 2$^+$ & 0$^+$ & 1$^-$ & 0$^-$ & 3$^+$ & 0$^+$ & 1$^+$ & 0$^-$ & 2$^-$ & 0$^+$ & 1$^-$ & 0$^-$ \\
\hline
\end{tabular}
\caption{The four-digit binary reflected Gray code (BRGC). The columns show words of the BRGC. The Hamming distance of the neighboring vectors is 1. The last row shows where the $j$th vector differs from the previous one and also shows in the right upper index if it is modified from zero to one ($+$) or from one to zero ($-$). (For $j=0$, comparison is made with the last word.) The index $j$ of the word or vector of the BRGC and $i$ of the digit of the BRGC are shown in the upper row and left column, respectively.}
\label{Hamming_table}
\end{table}

\begin{table} []
    \begin{tabular}{>{}l<{} *{19}{c}}
      \multicolumn{1}{l}{row index $i$} &&&&&&&\\\cline{1-1} 
      3 &&&&&&&&\colorBB 3&&\colorRR 3&&&&&\\
      2 &&&&&&&\colorBB 2&&\colorRR 3&&\colorBB 2&&&&\\
      1 &&&&&\colorBB 1&&\colorRR 2&&\colorBB 2&&\colorRR 2&&\colorBB 1&&\\
      0 &\colorBB 0&&\colorRR 1&&\colorBB 1&&\colorRR 1&&\colorBB 1&&\colorRR 1&&\colorBB 1&&\colorRR 1&&\colorBB 0&
    \end{tabular}
\caption{This table describes the number of consecutive zeros and ones in Table~\ref{Hamming_table} as the power of two. The exponent of the number of zeros is written in blue, and the exponent of the number of ones is written in red. We can easily observe that each row starts with $2^{i}$ zeros. When $i=3$ the $2^{i}$ zeros are followed by the same number of ones. When $i < 3$, the zeros are followed by $2^{i+1}$ consecutive ones and then again by zeros until the last sequence of $2^{i}$ zeros. This pattern can be described by Eq.~(\ref{Hamming_equation}).}
\label{Hamming_table_pattern}
\end{table}

It is also possible to figure out which component of the $j$th vector is different from the previous one. This can be calculated by ensuring the following condition to be satisfied:
\begin{equation}
    \exists Z \in \mathbf{Z}_0^+:\text{ } \frac{j}{2^i} = 2Z +1,
    \label{Hamming_equation_change}
\end{equation} 
that is, $j/2^i$ is an odd number; note that this defines $i$ uniquely. Alternatively, we can actually say that $i$ is the power of $2$ appearing in the prime factorization of $j$. Eq.~(\ref{Hamming_equation_change}) is proved in \ref{Appendix_2}.

Once the relevant position $i$ and the corresponding value $G_{i,j}$ of the $j$th word of the BRGC is known, the vector $\mathbf{m}$ can be updated as
\begin{equation}
    m^{(j)}_y = m^{(j-1)}_y + 2(2G_{i,j}-1)M_{iy} \equiv m^{(j-1)}_y + 2a^{(j)}_iM_{iy}
    \label{eq:update}
\end{equation} 
for all $y$; the corresponding $L^*_1$ Bell value will be obtained by taking the Manhattan norm of $\mathbf{m}^{(j)}$.

\subsubsection{Parallelization and further technical aspects}\label{sec:technical1}
When we use a GPU or some other distributed computing resource to calculate the $L_1$ norm, our implementation intends to distribute the tasks equally between the execution units or threads. In particular, we divide the core of the computational duty by distributing the possible strategy vectors along with the calculation of the corresponding $L^*_1$ Bell values among different threads in a nearly uniform way.

In order to do so in practice, we need to define an interval of $j$ for each thread, i.e., a minimal and a maximal $j$ for which the corresponding words in the BRGC should be calculated according to Eqs.~(\ref{Hamming_equation})-(\ref{Hamming_equation_change}). This is done in the following way. Let $C = 2^{n-1}$ be the number of all of the words to be considered, where $n$ is the length of the strategy vector (or, equally, the number of rows of the matrix), and $T$ be the number of threads. The indexing of the threads is provided by $t_j \in \{0, 1, \cdots, T - 1\}$. Note that $t_j$ is also a function associating a thread index to a given word index $j$. Let us furthermore define $J = \left\lfloor \frac{C}{T} \right\rfloor$, which is just the number of words any given thread should consider provided $C$ is divisible by $T$. To take into account cases when $C$ is not divisible by $T$, the remainder $R = C \mod T$ will also be needed. The minimal ($j_\mathrm{min}$) and maximal ($j_\mathrm{max}$) word indices corresponding to a given thread with index $t_j$ are calculated as follows.

\begin{algorithm}
\caption{Algorithm distributing the tasks equally among the threads.}\label{algorithm_min_max}
    \begin{algorithmic}
        \State $j_\mathrm{min} \gets t_j \times J$
        \State $j_\mathrm{max} \gets \left(t_j + 1\right) \times J - 1$
        \If{$t_j \le R$}
            \State $j_\mathrm{min} \gets j_\mathrm{min} + t_j$
        \Else
            \State $j_\mathrm{min} \gets j_\mathrm{min} + R$
        \EndIf
        \If{$t_j < R$}
            \State $j_\mathrm{max} \gets j_\mathrm{max} + t_j +1$
        \Else
            \State $j_\mathrm{max} \gets j_\mathrm{max} + R$
        \EndIf
    \end{algorithmic}
\end{algorithm}

After the $j_\mathrm{min}$ and $j_\mathrm{max}$ values are defined for each thread, the $j_\mathrm{min}$th word of the BRGC needs to be calculated with, e.g., Eq.~(\ref{Hamming_equation}) due to parallelization. After the $j_\mathrm{min}$th BRGC word is determined, the code calculates (and stores) the corresponding vector $\mathbf{m}$ by a vector-matrix multiplication, and obtains the corresponding $L^*_1$ Bell value as the Manhattan norm of this vector. The consecutive step is to determine the entry where the next BRGC word is changed, for which Eq.~(\ref{Hamming_equation_change}) is used. The vector $\mathbf{m}$ corresponding to that word is then calculated with the update rule of Eq.~(\ref{eq:update}); its Manhattan norm will provide the corresponding $L^*_1$ Bell value. This is just what is illustrated through Eqs.~(\ref{L1_calculate_simplified_method_second})-(\ref{L1_calculate_simplified_method_third}) and can be continued until $j_\mathrm{max}$ is reached. During the calculations, each thread keeps track of the maximal $L^*_1$ Bell value it has encountered along with the corresponding strategy vector $\mathbf{a}$. Once all of the threads have reached $j_\mathrm{max}$, the thread-wise maximal $L^*_1$ Bell values are compared with each other to select the global maximum.

Utilizing a GPU for parallelization has a peculiar aspect. We first need to point out that we can divide the $h$-digit BRGC, with $2^h$ words, into $2^l$ groups, where $1 \le l < h$. Let the groups be indexed by $g \in \{0, 1, \cdots, 2^l-1\}$. One group consists of $2^{h-l}$ words of the BRGC, and the global index $j$ of any of these words can be written as $j=g \cdot 2^{h-l}+k$, where $k \in \{0, 1, \cdots, 2^{h-l}-1\}$, as in Table~\ref{Hamming_table}. Note that in all different groups for a given $k > 0$, the change in the BRGC occurs at the same $i$th digit;
see \ref{Appendix_3} for a proof. This observation implies the following: if the number of threads is a power of $2$, then after the initial step, all of the threads will perform the same task just with different data.
This is beneficial for the calculation of the $L_1$ norm, since GPUs are devices applying the single instruction, multiple threads (SIMT) model. Namely, the physical execution units (cores) in a GPU are grouped into warps which usually consist of 32 cores. Within a warp, the threads perform the same instruction. If the cores within a warp need to perform different tasks, the different cores wait for each other until the given execution is performed for all of the threads. This is called branch divergence, and in our case, it can occur if the entry of the change in the BRGC words is found at different points of execution by the different cores in a given warp. Therefore, the number of threads should be specified as a power of $2$ even if the number of physical processing units does not have this property, cf. section~\ref{sec:performance}. This recommendation, however, does not apply to the OpenMP and MPI implementations as they utilize CPUs which are designed according to the multiple instruction, multiple data (MIMD) model.

We must point out that the simplest way to ensure our iterative method to work is to represent numbers as integers, as addition is not associative for floating-point numbers, and errors can grow large for a number of addition operations like that in our case. Using integers opens the opportunity to calculate the powers of two by bit shifting (instead of using the built-in power function, for instance), which significantly increases the performance of the code, especially with regard to Eqs.~(\ref{Hamming_equation})-(\ref{Hamming_equation_change}).

Considering the way the algorithm determines the $j$th word of the Gray code, by Eqs.~(\ref{Hamming_equation}) and (\ref{Hamming_equation_change}), it gives a constraint on the length of the Gray code the program can handle. This constraint is determined by the number of bits the integer is represented on a computer. If the integer is represented on $B$ bits, the program can determine Gray codes of length $B-2$ having $2^{B-2}$ words. As the variables are defined as \textit{unsigned long long int} which is stored in 64 bits, the code can calculate Gray codes of length 62. As the code considers one row with a fixed sign, the strategy vector can be of length $B-1$, which is 63 for 64-bit unsigned integers. Since the program transposes the input matrix if the number of columns exceeds the number of rows, the eventual constraint is that the number of rows and the number of columns of the input matrix cannot be more than 63 simultaneously.

\subsection{Special notes for calculating $L_{marg}$}
In case one intends to calculate the local bound of Bell expressions including marginals (see Eq.~\ref{LMarg}), the previous statements for preprocessing the witness matrix remain valid with the following constraints:
\begin{itemize}
    \item Trivially, when all entries of a given row or column are zero except for the first row or column; such a row or column can be removed. The first row or column can be removed when both the entries of the first row and the entries of the first column are exclusively zero. In this case the program will calculate the $L_1$ norm of the reduced matrix. 
    \item When a row or column of the matrix is a constant multiple of another row or column except for the first row or column. In this case, the two rows or columns can be replaced by a single one by adding or subtracting the two, depending on the sign of the constant multiplier.
\end{itemize}
According to Eq.~\ref{LMarg}, the rest of the computation can be performed as described above except that the fixed entry of the strategy vector must be chosen to be the first one and the first column needs to be treated separately from the rest.

\subsection{$L_\num$ norm, $\num\geq 2$}
In the following, we will discuss the efficient way of determining the $L_\num$ norm of a matrix when $\num > 1$, see Eq.~\ref{L_o_norm}. Our computation method is similar to that of the $L_1$ norm including simplifications and speedups (with the replacement of $L^*_1$ Bell values by $L^*_\num$ witness values), and we provide a discussion below that emphasizes the differences.

\subsubsection{Preprocessing the matrix}
An important difference with respect to the case of the $L_1$ norm is that transposition is not allowed, as it does not leave the $L_\num$ norm invariant for $\num > 1$. This introduces a limitation to an otherwise similar opportunity for reducing the number of rows; however, a new opportunity arises for cloumns. In total:
\begin{itemize}
  \item Any row or column having only zero entries can still be removed.
  \item When a column of the matrix is a constant multiple of another column, these columns can be replaced by a single one as for $L_1$. However, a similar replacement of two rows remains allowed only if the constant multiplier is positive.
  \item As a novelty, unification of two columns of the matrix having only non-negative or non-positive entries each, by addition in absolute value, becomes allowed.
\end{itemize}
See \ref{AppB} and \ref{AppA} for proofs.

\subsubsection{Algorithmic speedups}
It is very much intuitive from the grouping picture that one entry of the strategy vector can be assigned a fixed value, which entails $\num^{n-1}$ different strategy vectors to be considered (this indicates the multiplicity of an optimal strategy vector for $\num > 1$ as well).

The iterative method can be carried over from the $L_1$ case by utilizing $\num$-ary Gray codes \cite{SankarPSISPCS2004, HerterTCS2018}, which are series of vectors (words) the entries (digits) of which can take $\num$ possible values with neighboring words differing in exactly one digit. Note that the $\num=2$ Gray code is the same binary Gray code as for the $L_1$ norm. In this case, no conversion is needed between strategy vectors $\mathbf{a}$ and words of the Gray code, as the possible values of their entries are the same. Such a series of words, the ternary reflected Gray code (TRGC) in particular, is depicted in Table~\ref{Hamming_table_L3}. In the general case, the $\num$-ary reflected Gray code is defined by the following equations:
\begin{align}
	S &= (0, 1, \cdots, \num-1, \num-1, \cdots, 1, 0)\label{N_ary_G2}\\
    I_{i,j} &= \left\lfloor\frac{j}{\num^{i}}\right\rfloor \mod(2\num)\label{N_ary_G1}\\
    G_{i,j} &= S_{I_{i,j}}.\label{N_ary_G3}
\end{align}
$S$ is an auxiliary vector here, helping describe the $\num$-ary reflected Gray code. By Eq.~\ref{N_ary_G3}, we select its $I_{i,j}$th element (with indexing starting from zero) to be the $i$th digit of the $j$th word in the Gray code, $G_{i,j}$. Note that using Eqs.~(\ref{N_ary_G1})-(\ref{N_ary_G3}) is less efficient than using Eq.~(\ref{Hamming_equation}) for the BRGC. For this reason, we keep utilizing Eq.~(\ref{Hamming_equation}) for $\num=2$; however, we use Eqs.~(\ref{N_ary_G1})-(\ref{N_ary_G3}) otherwise for determining $G_{i,j}$ as a function of $j$ and $i$.

The digit that has changed in a given, $j$th, word of the $\num$-ary reflected Gray code with repect to the previous word (specified in the bottom row in Table~\ref{Hamming_table_L3} for the TRGC) can be found using the following condition:
\begin{equation}
    \max \left\{i: j \mod\num^i = 0 \right\}.
    \label{d_aryGrayChange}
\end{equation}
This formula can be evidently seen from the definition of the $\num$-ary reflected Gray code, Eqs.~(\ref{N_ary_G1})-(\ref{N_ary_G3}). 
Furthermore, it is equivalent to the expression
\begin{equation}
    \min \left\{i: \left\lfloor\frac{j}{\num^i}\right\rfloor \mod \num \neq 0 \right\},
    \label{d_aryGrayChange_1}
\end{equation}
see a proof in \ref{Appendix_4}. Eq.~(\ref{d_aryGrayChange_1}) is computationally more convenient than Eq.~(\ref{d_aryGrayChange}).

Computing $G_{i,j}$ as a function of $j$ and $i$ and finding the position $i$ of the change in the Gray code for consecutive values of $j$ make utilizing the iterative method possible for $\num > 1$. However, between consecutive iterations, all of the vectors $\mathbf{m}_a$, $a\in\{0, 1, \cdots, \num-1\}$, as defined in Eq.~\ref{eq:ma}, need to be stored, and updated upon each iteration as
\begin{align}
    \left(m^{(j)}_{G_{i,j-1}}\right)_y = \left(m^{(j-1)}_{G_{i,j-1}}\right)_y - M_{iy} , \\
    \left(m^{(j)}_{G_{i,j}}\right)_y = \left(m^{(j-1)}_{G_{i,j}}\right)_y + M_{iy}
    \label{eq:update2}
\end{align}
for all $y$.

\subsubsection{Technical considerations}
The design of parallelization is exactly the same as for the $L_1$ norm. We emphasize that branch divergence can be avoided in principle by specifying the total number of threads as a power of $\num$. However, we will see in section~\ref{sec:performance} that this is presumably not the full picture about performance.
\begin{table} [h]
\setlength{\tabcolsep}{2.0pt}
\renewcommand{\arraystretch}{1.2}
\begin{tabular}{rccccccccccccccccccccccccccc}
\multicolumn{27}{c}{$j$th word} \\

 \vline & 0 & 1 & 2 & 3 & 4 & 5 & 6 & 7 & 8 & 9 & 10 & 11 & 12 & 13 & 14 & 15 & 16 & 17 & 18 & 19 & 20 & 21 & 22 & 23 & 24 & 25 & 26 \\
\hline
\multirow{3}{0.9em}{\rotatebox[origin=c]{90}{$i$th digit}} 2 \vline & 0 & 0 & 0 & 0 & 0 & 0 & 0 & 0 & 0 & 1 & 1 & 1 & 1 & 1 & 1 & 1 & 1 & 1 & 2 & 2 & 2 & 2 & 2 & 2 & 2 & 2 & 2 \\
 1 \vline & 0 & 0 & 0 & 1 & 1 & 1 & 2 & 2 & 2 & 2 & 2 & 2 & 1 & 1 & 1 & 0 & 0 & 0 & 0 & 0 & 0 & 1 & 1 & 1 & 2 & 2 & 2 \\
 0 \vline & 0 & 1 & 2 & 2 & 1 & 0 & 0 & 1 & 2 & 2 & 1 & 0 & 0 & 1 & 2 & 2 & 1 & 0 & 0 & 1 & 2 & 2 & 1 & 0 & 0 & 1 & 2 \\
\hline
& & 0 & 0 & 1 & 0 & 0 & 1 & 0 & 0 & 2 & 0 & 0 & 1 & 0 & 0 & 1 & 0 & 0 & 2 & 0 & 0 & 1 & 0 & 0 & 1 & 0 & 0 \\
\hline
\end{tabular}
\caption{The three-digit ternary reflected Gray code (TRGC). The column vectors consist of three possible numbers, $\{0,1,2\}$, and represent strategies. The Hamming distance of the neighboring vectors is 1. The last row shows where the $j$th vector is different from the previous one. The upper row displays the index $j$ of the vector, and the left column displays the index $i$ of the digit of the TRGC.}
\label{Hamming_table_L3}
\end{table}

For a $\num$-ary Gray code we also have a constraint on the length of the Gray code the computer can calculate. Considering Eqs.~(\ref{N_ary_G1})-(\ref{N_ary_G3}), the maximal length of the $\num$-ary Gray code the computer can handle is given by
\begin{equation}
    N_\mathrm{max} = \left\lfloor\frac{B}{ \log_2(\num)} \right\rfloor,
\end{equation}
where $B$ is once more the number of bits the computer can represent an integer on, and $\num$ is the very order of the $\num$-ary Gray code. As the code fixes one value of the strategy vector, the dimension of the strategy vector as well as the maximal number of rows in the input matrix can be $n_\mathrm{max} = N_\mathrm{max} + 1$.

\section{A simple analysis of the speedup}

As we are claiming that our speedup relies on several factors, it is natural to elaborate on the issue and perform some test for verification. Parallelization itself is an obvious source of speedup, and we will illustrate here that our solution provides more both in theory and practice. We will concentrate on the $L_1$ norm of $n \times n$ square matrices as an example, and our reference will be the algorithm used by Brierley~et~al.~\cite{BrierleyArxiv2017} which is perhaps the most naive implementation of computing this norm.

In particular, a brute force computation must consider approximately $2^n$ different strategy vectors, which is the main contribution in the scaling of the computation time with $n$. For each strategy vector, Brierley~et~al. computed the Bell value from ``scratch'', which requires performing $n^2$ operations (a matrix-matrix multiplication in their code). However, an iterative algorithm like ours does this at the beginning only, and updates the Bell value for each new strategy vector by accessing a single row of the matrix in question, resulting in $n$ operations in each such iteration. Hence, the naive implementation is expected to scale with $n$ as $n^2 2^n$, whereas an iterative one as $n 2^n$.

Our particular code uses BRGC to achieve an iterative implementation. In order to compare its performance with the algorithm used by Brierley~et~al., we implemented the latter in C to eliminate the difference in performance between programming languages. As we are interested here in the algorithmic speedup as opposed to parallelization, we ran our programs utilizing a single thread; concretely, on an Intel(R) Xeon(R) Gold 6448Y processor after compiling it by the GNU Compiler Collection with no optimization (\texttt{-O0}). We simply measured the elapsed time for the two methods with matrices of different sizes; to reduce random contributions from unrelated system load, we actually performed the measurement three times and took the minimum.

\begin{figure}[h]
\centering\includegraphics[width=14.0cm]{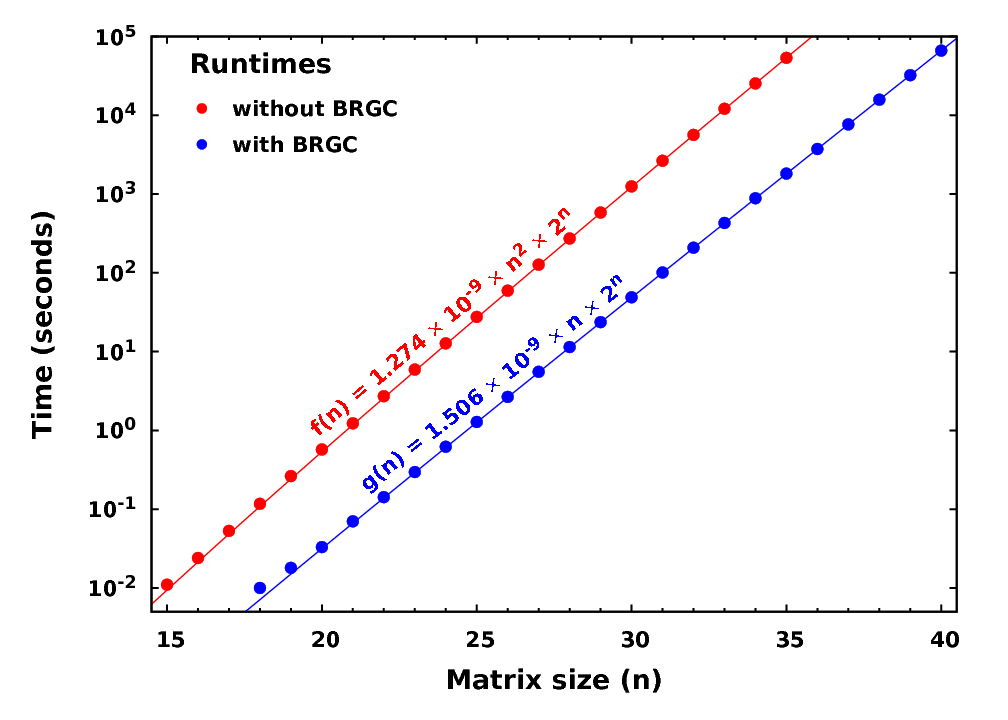}
\caption{Execution times for computing the $L_1$ norm for matrices with different sizes ($n \times n$) using a single core of an Intel(R) Xeon(R) Gold 6448Y processor. The codes were written in C and compiled using the GNU Compiler Collection with no optimization (\texttt{-O0}). The blue dots depict the execution times for the code using our algorithm, meanwhile, the red dots depict the execution times for a naive code (see text). Least-squares fits are indicated in the plot.} \label{times}
\end{figure}

The measurement results can be seen in Fig.~\ref{times}. The red and blue dots are depicting the execution times as the function of the size $n$ of the matrices using the algorithm of Brierly et al. and ours, respectively. Functions of the form $f(n) = A \times n^2 \times 2^n$ and $g(n) = B \times n \times 2^n$, respectively, fit very well to the measurement results, where the constants $A$ and $B$ have been fitted according to least squares. This confirms our advantage of a factor of $n$ in terms of scaling; furthermore, the prefactor $B$ turns out to be similar to $A$, indicating that our algorithm indeed has a major algorithmic advantage for matrices of relevant sizes.

Our results and some further experience (not shown) also suggest that administrative costs at the beginning or during the bulk of the computation are negligible in comparison with the full execution time. In fact, this is why we did not include an additive constant into functions $f(n)$ and $g(n)$. As an example for administrative costs during the bulk, calculating the next strategy vector does have a computational cost. However, in the case of Brierley~et~al., the calculation of the strategy vector out of a non-negative number requires bit operations, which is very efficient. On the other hand, our iterative algorithm determines the index where the next BRGC word has changed. One can notice from Eq.~\ref{Hamming_equation} and Table~\ref{Hamming_table} that $1/2$ of the time the change between the consecutive BRGC words occur at the 0th digit, $1/4$ of the time it happens at the 1st digit, $1/8$ of the time it happens at the 2nd digit and so on. If we consider an arbitrarily chosen word of an $n$-digit BRGC, the expected value of the times we need to calculate Eq.~\ref{Hamming_equation_change} is $\sum_{i=1}^{n} \frac{i}{2^{i}}$. As $\sum_{i=1}^{\infty} \frac{i}{2^{i}} = 2$ and $n<\infty$, we conclude that computing the next BRGC word in our algorithm is not a task that could strongly influence execution times and their scaling with $n$. We also need to mention that Eq.~\ref{Hamming_equation_change} is implemented with bit operations as well.

\section{Efficiency of the computations at different grid and block sizes}\label{sec:performance}
\begin{figure}
\vspace{-3cm}
\centering\includegraphics[width=14.0cm]{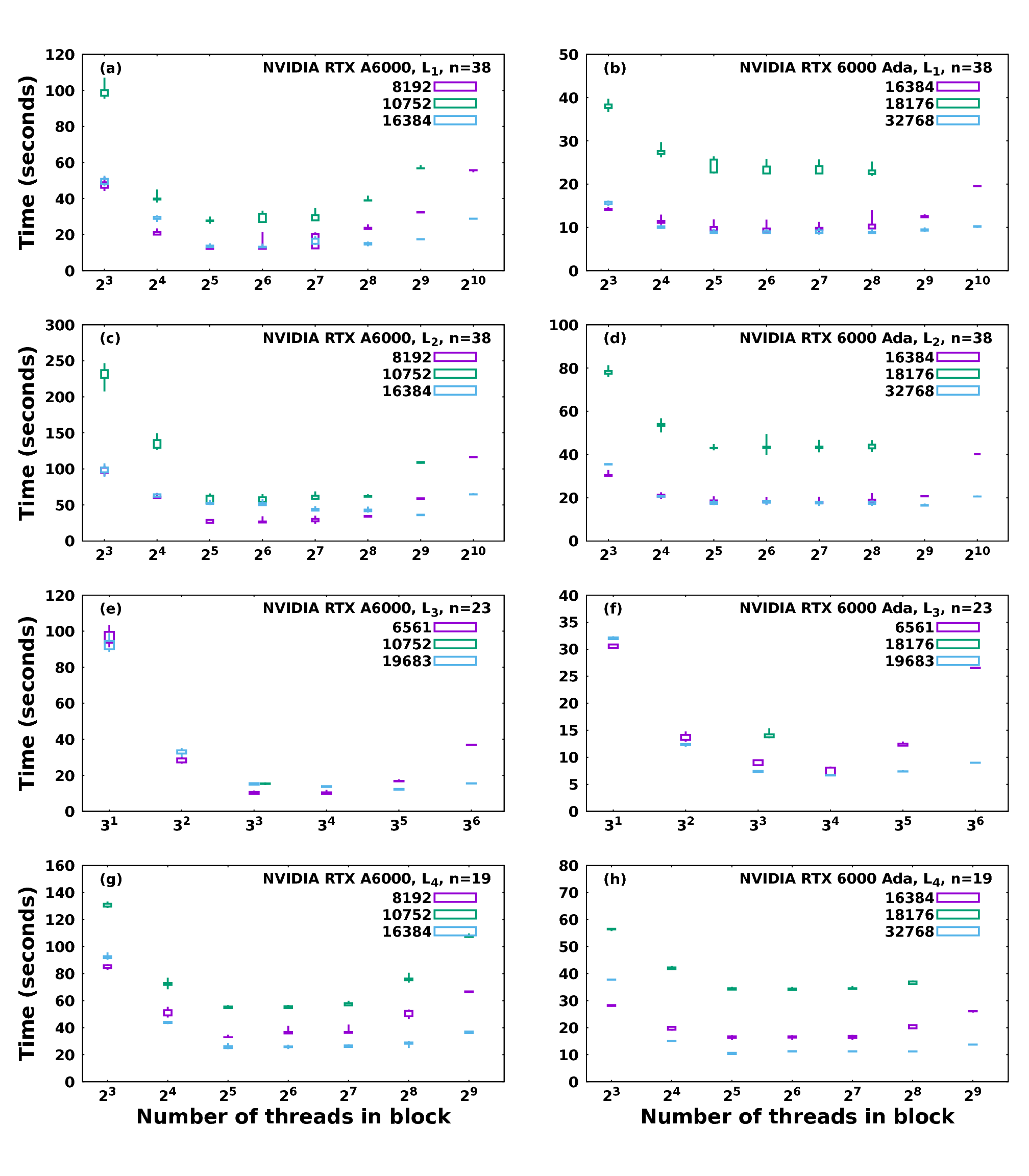}
\caption{Execution times for computing $L$ norms of different order for $n \times n$ matrices, as a function of the number of threads in a block along with keeping the total number of threads constant (for markers plotted with the same color and as indicated in the legend). In the left and right columns, execution times for NVIDIA RTX A6000 and NVIDIA RTX 6000 Ada GPUs are depicted, respectively. Different rows correspond to different orders of the $L$ norm, ordered as $L_1$ to $L_4$ from top to bottom. The size $n$ of the matrix is indicated within each subplot. Box plotting is used to display the minimal execution time, the first quartile, the second quartile, and the maximal execution time for a given matrix among 50 different executions.} \label{execution_times}
\end{figure}
As we mentioned earlier, the grid size, namely the number of blocks, and the block size, namely the number of threads in a block, can influence the performance of the code. We have performed a dedicated investigation with two different GPUs, namely an NVIDIA RTX A6000 and an NVIDIA RTX 6000 Ada having 10752 and 18176 CUDA cores, respectively.

We studied the performance of the CUDA code for the $L_1,\ldots,L_4$ norms with square ($n \times n$) matrices. As the execution time increases with the order of the $L$ norm of a matrix, we used smaller matrices for calculating higher order $L$ norms. In order to take the variation of the execution times between repeated calls into account, we performed the computation 50 times for a given grid and block size, and we represent the distibution of the results by box plots in Fig.~\ref{execution_times}. Markers having the same color depict execution times for cases when the product of the grid size and the block size is kept constant, i.e., for which the total number of threads (TNOT) is the same.

As it can be seen from Fig.~\ref{execution_times}, execution is not the fastest when TNOT is the same as the number of CUDA cores in the GPU. Instead, performance is optimal when TNOT is a power of $\num$ or at least ``closely related'' (see later). Furthermore, efficiency is influenced by the block size to some extent. Although a considerably large neighborhood of the optimum is relatively flat as the function of block size, this effect becomes really important beyond certain amount of deviation from the optimum, especially in the direction of small block sizes (which can be regarded as intuitive).

In particular, the shortest execution times for $L_1$ (see Figs.~\ref{execution_times}.~(a)-(b)) were measured on both GPUs at a block size of 32 or 64, and choosing TNOT to be a power of $2$ provides an important improvement with respect to the number of CUDA cores, which would be easy to explain by avoidance of branch divergence, cf. section~\ref{sec:technical1}. The conclusions are similar for $L_2$ (Figs.~\ref{execution_times}.~(c)-(d)), but we need to mention that the performance for the RTX~A6000 GPU (see Fig.~\ref{execution_times}.~(c)) is conspicuously better for a TNOT value of 8192 in comparison with 16384 (as opposed to practically no difference for other cases discussed so far). In the case of $L_3$ (Figs.~\ref{execution_times}.~(e)-(f)), TNOT needs to be chosen as a power of $3$ for optimal performance, whether exceeding the number of CUDA cores or not (the differences appear to remain minor around the optimal block size). Interestingly, $L_4$ (Figs.~\ref{execution_times}.~(g)-(h)) features an optimum when TNOT just exceeds the CUDA cores and is the power of $2$ (with a considerable advantage with respect to the previous power of $2$); this means 32768 for the RTX~6000~Ada GPU (see Fig.~\ref{execution_times}.~(g)), which is \emph{not} a power of $4$. Branch divergence is thus not fully avoided in terms of the main computational task in this case, so that it probably cannot provide a full explanation for the optimal choice of TNOT.

The presented results indicate that the execution time is influenced, besides the order of the $L$ norm, by the graphics card, the block size, and TNOT in the first place. We did not find a general rule to determine the optimal block and grid size, only a few guiding principles. Note that it is possible and also mandatory to provide these parameters to the program before execution. Therefore, it is possible and we highly recommend to test the available GPU for the optimal parameters. A reasonable strategy is to do so for a relatively smaller matrix and then choose the parameters for more demanding calculations based on the results.

\section{Final remarks}
In this article, we have presented a method to efficiently calculate the $L_\num$ norm of a matrix, for any given $\num \geq 1$, in a parallel environment. Apart from their significance in the evaluation of the classical bounds on Bell inequalities or prepare-and-measure witnesses, these norms are also of interest in the field of communication complexity, the Grothendieck constant, or in graph theory. In order to be able to consider all two-outcome bipartite Bell inequalities, our program can also calculate the local bound of Bell expressions including marginals. As a result of parallelization (with special emphasis on GPUs) and the use of special mathematical and programming techniques, our implementation has achieved a significant speedup factor in comparison with previous computations performed on a single CPU core (see e.g.~related numerics-focused works under Refs.~\cite{GrucaPRA2010, GondzioJCompApplMath2014, BrierleyArxiv2017, Arajo2020} on computing the $L_1$ norm). We believe that our primitives can also be used as an improved simulation of the oracle in the optimization algorithms of Refs.~\cite{BrierleyArxiv2017, MontinaEntropy2019, Designolle2023, Designolle2024}, thus further boosting their performance.

\section*{Declaration of Competing Interest}
The authors declare that they have no known competing financial interests or personal relationships that could have appeared to influence the work reported in this paper.

\section*{Acknowledgments}
We thank Tamás Vértesi, Kai-Siang Chen and Hsin-Yu Hsu for valuable input and fruitful discussions. The authors thank Csaba Hruska for providing technical assistance in the GPU computations. This paper was supported by the János Bolyai Research Scholarship of the Hungarian Academy of Sciences, as well as the EU (CHIST-ERA MoDIC) and the National Research, Development and Innovation Office NKFIH (No.~2023-1.2.1-ERA\_NET-2023-00009 and No.~K145927). G.D. acknowledges support from the EU (European Social Fund Plus) and the Government of the Balearic Islands through a Vicen\c{c} Mut postdoctoral fellowship (No. PD-035-2023).

\appendix
\section{Proving the invariance of $L_\num$ under uniting specific columns or rows that are constant multiples of each other}
\label{AppB}

When a column or a row of a matrix $M$ is a constant multiple of another column or row of the given matrix, some simple rules for matrix reduction can be applied during the calculation of the $L_\num$ norms:
\begin{description}
\item [A1.]  For $d=1$, two such columns or rows can be added if the constant is positive, and can be subtracted (in arbitrary order) if the constant is negative.
\item [A2.] For $d\geq2$, the same is true as A1 for two such columns.
\item [A3.] For $d\geq2$, two such rows can only be added, and only if the constant is positive.
\end{description}

For proving these statements, we introduce the notation $M'$ for the reduced matrix.

\begin{description}

\item [Proof of A1.] Suppose that $M_{xy''}=cM_{xy'}$ for all $x$, where $y' \neq y''$. Then we have
\begin{align}
    L^*_1(M,\mathbf{a})
    &= \left| \sum_{x} M_{x y'} a_x \right| + \left| \sum_{x} M_{xy''} a_x \right| + \sum_{y \neq y', y''} \left| \sum_{x} M_{xy} a_x \right| \nonumber\\
    &= \left| \sum_{x} M_{x y'} a_x \right| + \left| c \right|  \left| \sum_{x} M_{xy'} a_x \right| + \sum_{y \neq y', y''} \left| \sum_{x} M_{xy} a_x \right| \nonumber\\
    &= (1+\left| c \right|)  \left| \sum_{x} M_{xy'} a_x \right| + \sum_{y \neq y', y''} \left| \sum_{x} M_{xy} a_x \right| \nonumber\\
     &= \left| \sum_{x} (1+\left| c \right|)  M_{xy'} a_x \right| + \sum_{y \neq y', y''} \left| \sum_{x} M_{xy} a_x \right|  = L^*_1(M',\mathbf{a}).
    \label{L1_reduce_1}
\end{align}
Note that an arbitrary column can be multiplied by $-1$ without affecting the $L_1$ norm, which ensures consistency with the freedom to choose an arbitrary order for subtraction. Since transposing the matrix leaves $L_1$ unchanged, our proof also covers the case of rows besides that of columns.

\item [Proof of A2.] In a similar fashion to the above proof, we can write the following chain of equations:
\begin{align}
    L^*_d(M,\mathbf{a})
     &=\sum_a \left( \left| \sum_{x: a_x = a} M_{x y'} \right| + \left| c  \right| \left| \sum_{x: a_x = a} M_{xy'} \right| +  \sum_{y \neq y', y''} \left|\sum_{x: a_x = a} M_{xy} \right| \right) \nonumber\\
      &= \sum_a \left( \left| \sum_{x: a_x = a} (1+\left| c  \right|) M_{xy'} \right| +  \sum_{y \neq y', y''} \left|\sum_{x: a_x = a} M_{xy} \right| \right) \nonumber\\ 
      &= L^*_d(M',\mathbf{a}),
    \label{Ld_reduce_cols}
\end{align} 
The same considerations apply to the order of subtraction.

\item [Proof of A3.] In this case we start from Eq.~(\ref{L2Mv2}), where $L_\num$ is defined as 
\begin{align}
    L_\num(M)
    &=\max_{\substack{\{a_x=0,+1,\cdots, \num-1\} \\ \{b_y^a = \pm1\}}} \left(\sum_{a} \sum_{x: a_x=a}\sum_{y} M_{xy}b^a_y \right) \nonumber\\
    &=\max_{\substack{\{b_y^a = \pm1\}}} \sum_{x} \max\left(\sum_{y} M_{xy}b^0_y,\sum_{y} M_{xy}b^1_y , \dots , \sum_{y} M_{xy}b^{\num-1}_y\right).
    \label{L_reduce_1}
\end{align}
Suppose that $M_{x''y} = c M_{x'y}$ for all $y$, where $c > 0$ and $x' \neq x''$. In this case we can write:
\begin{equation}
    \sum_y M_{x'' y} b_y^a = c \sum_y M_{x' y} b_y^a 
    \label{eq:positive_constant}
\end{equation}
for all $a\in\{0,\ldots,d-1\}$. We then separate the above expression for $L_\num$ according to the value of $x$ as follows:
\begin{align}
    L_\num(M)
    =&\max_{\{b_y^a = \pm1\}} \left[ \sum_{x\in\{x', x''\}} \max \left( \sum_y M_{xy} b_y^0, \dots ,\sum_y M_{xy} b_y^{d-1} \right) 
    +\right.\nonumber\\ 
    &\quad + \left. \sum_{x \neq x', x''} \max \left( \sum_y M_{xy} b_y^0, \dots , \sum_y M_{xy} b_y^{d-1} \right) \right]
    \label{L_reduce_2}
\end{align}
and according to Eq.~(\ref{eq:positive_constant}) we can write
\begin{align}
    &\sum_{x\in\{x',x''\}} \max \left( \sum_y M_{xy} b_y^0,\dots , \sum_y M_{xy} b_y^{d-1} \right) =\nonumber\\ &\left( c+1 \right) \cdot \max \left( \sum_y M_{x'y} b_y^0, \dots ,\sum_y M_{x'y} b_y^{d-1} \right),
    \label{L_reduce_3}
\end{align}
from which it directly follows that rows $x'$ and $x''$ can be added.

\end{description}

\section{Proving the invariance of $L_\num$, $d\geq2$, when adding specific columns}
\label{AppA}

We will prove here that any two columns of a matrix $M$ that each have elements of uniform sign can be replaced by a single column composed as the sum of the absolute values of the elements in these two columns.

Let $y_1$ and $y_2$ be the indices of two such columns. By definition, the $L^*_\num$ witness value for an arbitrary but given strategy vector $\mathbf{a}$ can be expressed as
\begin{align}
	L^*_d(M,\mathbf{a})
	&= \sum_{a} \left( \sum_{y\neq y_1,y_2} \left|\sum_{x:a_x=a} M_{xy}\right| + \left|\sum_{x:a_x=a} M_{xy_1}\right| + \left|\sum_{x:a_x=a} M_{xy_2}\right| \right) .
\end{align}
We now take advantage of the property that $\sgn M_{x_1y} = \sgn M_{x_2y}$ for all $x_1,x_2$ for both $y=y_1$ and $y=y_2$ (note that the sign of the elements may be different in the two columns):
\begin{align}
	L^*_d(M,\mathbf{a}) &= \sum_{a} \left( \sum_{y\neq y_1,y_2} \left|\sum_{x:a_x=a} M_{xy}\right| + \sum_{x:a_x=a} \left|M_{xy_1}\right| + \sum_{x:a_x=a} \left|M_{xy_2}\right|) \right) \nonumber \\
	&= \sum_{a} \left( \sum_{y\neq y_1,y_2} \left|\sum_{x:a_x=a} M_{xy}\right| + \left| \sum_{x:a_x=a} \left(\left|M_{xy_1}\right| + \left|M_{xy_2}\right|\right) \right| \right) .
\end{align}
From this point, it is straightforward to recover the relevant $L^*_d$ witness value of the replacement matrix $M'$, the replacement column $y'$ of which is defined as
\begin{equation}
	M'_{xy'} = \left|M_{xy_1}\right| + \left|M_{xy_2}\right|
\end{equation}
for all $x$, and which has identical columns to $M$ otherwise. In particular,
\begin{align}
	L^*_d(M,\mathbf{a})
	&= \sum_{a} \left( \sum_{y\neq y_1,y_2} \left|\sum_{x:a_x=a} M_{xy}\right| + \left|\sum_{x:a_x=a} M'_{xy'}\right| \right) \nonumber \\
	&= \sum_{a} \sum_y \left|\sum_{x:a_x=a} M'_{xy}\right| 
	= L^*_d(M',\mathbf{a}) .
\end{align}
 
\section{Proof of Eq.~(\ref{Hamming_equation}) with mathematical induction} \label{Appendix_1}
Formula~(\ref{Hamming_equation}) can be proved by mathematical induction. For $n = 1$, our claim is trivial. Then assume that for the $n$-digit BRGC Eq.~(\ref{Hamming_equation}) holds, i.e.,
\begin{equation}
    G_{i,j} = \left\lfloor\frac{j + 2^{i}}{2^{i + 1}}\right\rfloor \mod 2; \text{ } i \in \{n-1, \ldots, 0\}; j \in \{0, \ldots, 2^n-1\}.
    \label{H}
\end{equation}
Using this assumption, we will show below that Eq.~(\ref{Hamming_equation}) also holds in the $(n+1)$-digit case.

The $(n+1)$-digit case is generated recursively from the $n$-digit case by reflecting the $n \times 2^n$ matrix $G$ and completing the first row with values $0$ for column indices $0 \le j \le 2^n-1$ and with values $1$ for $2^n \le j \le 2^{n+1}-1$, as it is depicted in Table~\ref{Hamming_nplus1}.

\begin{table} [h]
\renewcommand{\arraystretch}{1.0}
\begin{center}
\begin{tabular}{c|c}
0, $\cdots$, 0 & 1, $\cdots$, 1\\
\hline
$G$ & $G'$ \\
\end{tabular}
\end{center}
\caption{Generating the $(n+1)$-digit BRGC from the $n$-digit BRGC, which is denoted by $G$. $G'$ is obtained by reflecting $G$.}
\label{Hamming_nplus1}
\end{table}

Depending on the value of $j$, two cases can be distinguished:
\begin{description}
\item [Case 1]
For $j \le 2^n-1$, our statement is trivial using the definition of the BRGC. For $i \in \{n-1, \ldots, 0\}$, it follows evidently from the induction assumption Eq.~(\ref{H}). For $i = n$, then $\left\lfloor\frac{j + 2^{i}}{2^{i + 1}}\right\rfloor \mod 2 = 0$, since $\frac{1}{2}+\frac{j}{2^{n+1}} < 1$.
\item [Case 2]
For $2^n \le j \le 2^{n+1}-1$, the definition and assumption~(\ref{H}) imply
\begin{equation}
    G_{i,j} = \left\lfloor\frac{2^{n+1} -1 -j + 2^{i}}{2^{i + 1}}\right\rfloor \mod 2; \text{ } i \in \{ n-1, \ldots, 0 \},
    \label{Hnplus1}
\end{equation}
and therefore we will prove the equality
\begin{equation}
    \left\lfloor 2^{n-i} - \frac{j}{2^{i+1}} + \frac{1}{2} - \frac{1}{2^{i+1}} \right\rfloor \mod 2 = \left\lfloor\frac{j}{2^{i + 1}} + \frac{1}{2} \right\rfloor \mod 2.
    \label{Hnplus1_2}
\end{equation}
The right-hand side of Eq.~(\ref{Hnplus1_2}) can be replaced by the expression
\begin{equation}
    \left\lfloor 2^{n-i} +\frac{j}{2^{i + 1}} + \frac{1}{2} \right\rfloor \mod 2.
    \label{Hnplus1_3}
\end{equation}
Decomposing $\frac{j}{2^{i+1}}$ into integer and fraction parts $Z = \left\lfloor \frac{j}{2^{i + 1}}\right\rfloor$ and $F = \frac{j}{2^{i+1}}-Z \in [0,1)$, the equality to be verified finally takes the form
\begin{equation}
\begin{split}
    \left\lfloor 2^{n-i} - Z + \frac{1}{2} - \frac{1}{2^{i+1}} -F \right\rfloor \mod 2 =\\ \left\lfloor 2^{n-i} + Z + \frac{1}{2} + F \right\rfloor \mod 2, \label{Hnplus1_4}
\end{split}
\end{equation}
This equation holds because the parity of $\left\lfloor 2^{n-i} -Z + \frac{1}{2} \right\rfloor$ is the same as the parity of $\left\lfloor 2^{n-i} + Z + \frac{1}{2} \right\rfloor$ and $\frac{1}{2^{i+1}} \le \frac{1}{2}$; it is instructive to consider the cases $F \ge \frac{1}{2}$ and $F < \frac{1}{2}$ separately. Note that Eq.~(\ref{Hnplus1}), just proved, does not cover $i = n$; for this case, $\left\lfloor\frac{j + 2^{i}}{2^{i + 1}}\right\rfloor \mod 2 = 1$, since $\frac{1}{2} \le \frac{1}{2}+\frac{j}{2^{n+1}} \le 1-\frac{1}{2^{n+1}}$.
\end{description}

\section{Proof of Eq.~(\ref{Hamming_equation_change})} \label{Appendix_2}
Let us write $j+2^i$ in the following form:
\begin{equation}
    j+2^i = Z \cdot 2^{i+1} + R; \text{ } 0 \le R < 2^{i+1};\text{ } Z,R \in \mathbb{Z}_0^+.
    \label{Hamming_equation_zero}
\end{equation}
According to Eq.~(\ref{Hamming_equation}), the $j$th and $(j-1)$th column can be expressed as
\begin{equation}
    G_{i,j} = \left\lfloor\frac{j + 2^{i}}{2^{i + 1}}\right\rfloor \mod 2 = \left\lfloor Z + \frac{R}{2^{i + 1}}\right\rfloor \mod 2
\end{equation}
and
\begin{equation}
    G_{i,j-1} = \left\lfloor\frac{j-1 + 2^{i}}{2^{i + 1}}\right\rfloor \mod 2 = \left\lfloor Z + \frac{R}{2^{i + 1}}- \frac{1}{2^{i + 1}}\right\rfloor \mod 2, \label{Hamming_equation_minus1}
\end{equation}
respectively.

If $R = 0$, the expressions $\left\lfloor Z + \frac{R}{2^{i + 1}}\right\rfloor$ and $\left\lfloor Z + \frac{R}{2^{i + 1}}- \frac{1}{2^{i + 1}}\right\rfloor$ have different parities as a result of the inequality $0 < \frac{1}{2^{i+1}} \le \frac{1}{2}$. If $R \neq 0$, these expressions have the same parities, because $\frac{R}{2^{i + 1}} < 1$ and $0 \le \frac{R}{2^{i+1}}-\frac{1}{2^{i+1}} < 1$. This means that the change occurs when the following condition is satisfied:
\begin{equation}
    \left(j + 2^{i}\right) \mod 2^{i + 1} = 0,
    \label{App_cond1}
\end{equation}
which is equivalent to Eq.~(\ref{Hamming_equation_change}).

We note that Eq.~(\ref{Hamming_equation_change}) can also be derived from Eq.~(\ref{d_aryGrayChange}) for $d=2$.

\section{Proving that the changes in the $\num$-ary Gray code groups occur at the same digits} \label{Appendix_3}

The $h$-digit $\num$-ary Gray code can be divided into $\num^l$ groups indexed by $g \in \{0, 1, \ldots, \num^l-1\}$ for some positive integer $l$ such that we write the global word index $j$ as $j = g \cdot \num^{h-l}+k$, $k \in \{0, \ldots, \num^{h-l}-1\}$. We will prove that the change occurs at the same digit $i$ within every group for any given $k \in \{ 1, \ldots, \num^{h-l}-1 \}$.

According to Eq.~(\ref{d_aryGrayChange}), the change occurs at the digit with the maximal $i$ such that $j$ is divisible by $\num^{i}$. Since $k \neq 0$, $j$ is not divisible by $\num^{h-l}$. This means $j$ is not divisible by $\num^i$ if $i \ge h-l$. This implies $i < h-l$ for the digit $i$ of the change.

Let us suppose that there exist two different groups $g$ and $g'$ for which the changes occur at different digits $i$ for a given $k \neq 0$, that is
\begin{equation}
    g \cdot \num^{h-l} + k = Z \cdot \num^i \label{App_change1}\\
\end{equation}
and
\begin{equation}
    g' \cdot \num^{h-l} + k = Z' \cdot \num^i + R,\label{App_change2}
\end{equation}
where $Z, Z', R \in \mathbb{Z}_0^+$ and $0 < R < \num^{i}$. Subtracting these two equations we obtain the expression
\begin{equation}
    \left(g' - g\right) \cdot \num^{h-l}=(Z'-Z) \cdot \num^i + R ,
\end{equation}
which is true when $R=0$. This means for a given $k$ the values $j=g \cdot \num^{h-l} + k$; $g = 0,1,\cdots, \num^l-1$ are divisible with the same maximal power of $\num$. This means for a given $k>0$, the changes occur at the same row in the given words of the $d$-ary reflected Gray code which was to be proved.

\section{Proof of the equivalence of Eqs.~(\ref{d_aryGrayChange}) and (\ref{d_aryGrayChange_1})} \label{Appendix_4}
Let us define $i_1$ and $i_2$ as
\begin{equation}
    i_1 = \max \left\{i: j \mod \num^i = 0 \right\}\label{App_d_aryGrayChange}\\
\end{equation}
and
\begin{equation}
    i_2 = \min \left\{i: \left\lfloor\frac{j}{\num^i}\right\rfloor \mod \num \neq 0 \right\}. \label{App_d_aryGrayChange_1}
\end{equation}
In the following, we will prove in two steps that $i_1 = i_2$. In the first step, we prove that
\begin{equation}
    \left\lfloor\frac{j}{\num^{i_1}}\right\rfloor \mod \num \neq 0. \label{App_divisibility}
\end{equation}
According to the definition of $i_1$, $j$ is divisible with $\num^{i_1}$ that is $\exists Z \in \mathbb{Z}_0^+: j = Z \cdot \num^{i_1}$, but $j$ is not divisible with $\num^{i_1+1}$. Therefore $Z$ is not divisible with $\num$ and because $\left\lfloor \frac{j}{\num^{i_1}} \right\rfloor = Z$ the Eq.~(\ref{App_divisibility}) is proved to be true.\\
As a second step, we need to prove that $i_1$ is minimal out of those indices for which $\left\lfloor \frac{j}{\num^{i}} \right\rfloor \mod \num \neq 0$ is true. Let us suppose that 
\begin{equation}
    \exists i^{'} < i_1 : \left\lfloor \frac{j}{\num^{i^{'}}} \right\rfloor \mod \num \neq 0.\label{App_condition}
\end{equation}
As $i^{'} < i_1$, $j$ is divisible with $\num^{i'}$, and $\left\lfloor \frac{j}{\num^{i^{'}}} \right\rfloor = \frac{j}{\num^{i^{'}}}$ is divisible with $\num$ that contradicts Eq.~(\ref{App_condition}) and we proved that $i_1 = i_2$, i.e. Eq.~(\ref{d_aryGrayChange}) and Eq.~(\ref{d_aryGrayChange_1}) are equivalent.

\bibliographystyle{elsarticle-num}
\bibliography{sample.bib}

\begin{thebibliography}{10}
\expandafter\ifx\csname url\endcsname\relax
  \def\url#1{\texttt{#1}}\fi
\expandafter\ifx\csname urlprefix\endcsname\relax\def\urlprefix{URL }\fi
\expandafter\ifx\csname href\endcsname\relax
  \def\href#1#2{#2} \def\path#1{#1}\fi

\bibitem{BrierleyArxiv2017}
S.~Brierley, M.~Navascues, T.~Vertesi, {Convex separation from convex optimization for large-scale problems} (2017).
\newblock \href {http://arxiv.org/abs/1609.05011} {\path{arXiv:1609.05011}}.

\bibitem{Nielsen2010}
M.~A. Nielsen, I.~L. Chuang, {Quantum computation and quantum information}, {Cambridge university press}, 2010.

\bibitem{Scarani2019}
V.~Scarani, \href{http://dx.doi.org/10.1093/oso/9780198788416.001.0001}{{Bell Nonlocality}}, Oxford University PressOxford, 2019.
\newblock \href {https://doi.org/10.1093/oso/9780198788416.001.0001} {\path{doi:10.1093/oso/9780198788416.001.0001}}.
\newline\urlprefix\url{http://dx.doi.org/10.1093/oso/9780198788416.001.0001}

\bibitem{BrunnerPRL2008}
N.~Brunner, S.~Pironio, A.~Acin, N.~Gisin, A.~A. M\'ethot, V.~Scarani, \href{https://link.aps.org/doi/10.1103/PhysRevLett.100.210503}{{Testing the Dimension of Hilbert Spaces}}, Phys. Rev. Lett. 100 (2008) 210503.
\newblock \href {https://doi.org/10.1103/PhysRevLett.100.210503} {\path{doi:10.1103/PhysRevLett.100.210503}}.
\newline\urlprefix\url{https://link.aps.org/doi/10.1103/PhysRevLett.100.210503}

\bibitem{GallegoPRL2010}
R.~Gallego, N.~Brunner, C.~Hadley, A.~Ac\'{\i}n, \href{https://link.aps.org/doi/10.1103/PhysRevLett.105.230501}{{Device-Independent Tests of Classical and Quantum Dimensions}}, Phys. Rev. Lett. 105 (2010) 230501.
\newblock \href {https://doi.org/10.1103/PhysRevLett.105.230501} {\path{doi:10.1103/PhysRevLett.105.230501}}.
\newline\urlprefix\url{https://link.aps.org/doi/10.1103/PhysRevLett.105.230501}

\bibitem{PalPRA2008}
K.~F. P\'al, T.~V\'ertesi, \href{https://link.aps.org/doi/10.1103/PhysRevA.77.042105}{{Efficiency of higher-dimensional Hilbert spaces for the violation of Bell inequalities}}, Phys. Rev. A 77 (2008) 042105.
\newblock \href {https://doi.org/10.1103/PhysRevA.77.042105} {\path{doi:10.1103/PhysRevA.77.042105}}.
\newline\urlprefix\url{https://link.aps.org/doi/10.1103/PhysRevA.77.042105}

\bibitem{WehnerPRA2008}
S.~Wehner, M.~Christandl, A.~C. Doherty, \href{https://link.aps.org/doi/10.1103/PhysRevA.78.062112}{{Lower bound on the dimension of a quantum system given measured data}}, Phys. Rev. A 78 (2008) 062112.
\newblock \href {https://doi.org/10.1103/PhysRevA.78.062112} {\path{doi:10.1103/PhysRevA.78.062112}}.
\newline\urlprefix\url{https://link.aps.org/doi/10.1103/PhysRevA.78.062112}

\bibitem{WolfPRL2009}
M.~M. Wolf, D.~Perez-Garcia, \href{https://link.aps.org/doi/10.1103/PhysRevLett.102.190504}{{Assessing Quantum Dimensionality from Observable Dynamics}}, Phys. Rev. Lett. 102 (2009) 190504.
\newblock \href {https://doi.org/10.1103/PhysRevLett.102.190504} {\path{doi:10.1103/PhysRevLett.102.190504}}.
\newline\urlprefix\url{https://link.aps.org/doi/10.1103/PhysRevLett.102.190504}

\bibitem{Scarani2009}
V.~Scarani, H.~Bechmann-Pasquinucci, N.~J. Cerf, M.~Du\ifmmode~\check{s}\else \v{s}\fi{}ek, N.~L\"utkenhaus, M.~Peev, \href{https://link.aps.org/doi/10.1103/RevModPhys.81.1301}{{The security of practical quantum key distribution}}, Rev. Mod. Phys. 81 (2009) 1301--1350.
\newblock \href {https://doi.org/10.1103/RevModPhys.81.1301} {\path{doi:10.1103/RevModPhys.81.1301}}.
\newline\urlprefix\url{https://link.aps.org/doi/10.1103/RevModPhys.81.1301}

\bibitem{Woodhead2015}
E.~Woodhead, S.~Pironio, \href{https://link.aps.org/doi/10.1103/PhysRevLett.115.150501}{{Secrecy in Prepare-and-Measure Clauser-Horne-Shimony-Holt Tests with a Qubit Bound}}, Phys. Rev. Lett. 115 (2015) 150501.
\newblock \href {https://doi.org/10.1103/PhysRevLett.115.150501} {\path{doi:10.1103/PhysRevLett.115.150501}}.
\newline\urlprefix\url{https://link.aps.org/doi/10.1103/PhysRevLett.115.150501}

\bibitem{AcinNature2016}
A.~Acín, L.~Masanes, \href{http://dx.doi.org/10.1038/nature20119}{{Certified randomness in quantum physics}}, Nature 540~(7632) (2016) 213–219.
\newblock \href {https://doi.org/10.1038/nature20119} {\path{doi:10.1038/nature20119}}.
\newline\urlprefix\url{http://dx.doi.org/10.1038/nature20119}

\bibitem{LiPRA2012}
H.-W. Li, M.~Paw\l{}owski, Z.-Q. Yin, G.-C. Guo, Z.-F. Han, \href{https://link.aps.org/doi/10.1103/PhysRevA.85.052308}{{Semi-device-independent randomness certification using $n\ensuremath{\rightarrow}1$ quantum random access codes}}, Phys. Rev. A 85 (2012) 052308.
\newblock \href {https://doi.org/10.1103/PhysRevA.85.052308} {\path{doi:10.1103/PhysRevA.85.052308}}.
\newline\urlprefix\url{https://link.aps.org/doi/10.1103/PhysRevA.85.052308}

\bibitem{AmbainisJACM2002}
A.~Ambainis, A.~Nayak, A.~Ta-Shma, U.~Vazirani, \href{https://doi.org/10.1145/581771.581773}{{Dense quantum coding and quantum finite automata}}, J. ACM 49~(4) (2002) 496–511.
\newblock \href {https://doi.org/10.1145/581771.581773} {\path{doi:10.1145/581771.581773}}.
\newline\urlprefix\url{https://doi.org/10.1145/581771.581773}

\bibitem{ScaraniActPhysSlov2012}
V.~Scarani, {The device-independent outlook on quantum physics}, Acta Physica Slovaca 62~(4) (2012) 347--409.

\bibitem{BowlesPRA2015}
J.~Bowles, N.~Brunner, M.~Paw\l{}owski, \href{https://link.aps.org/doi/10.1103/PhysRevA.92.022351}{{Testing dimension and nonclassicality in communication networks}}, Phys. Rev. A 92 (2015) 022351.
\newblock \href {https://doi.org/10.1103/PhysRevA.92.022351} {\path{doi:10.1103/PhysRevA.92.022351}}.
\newline\urlprefix\url{https://link.aps.org/doi/10.1103/PhysRevA.92.022351}

\bibitem{HendrychNatPhys2012}
M.~Hendrych, R.~Gallego, M.~Mi{\v{c}}uda, N.~Brunner, A.~Ac{\'i}n, J.~P. Torres, \href{https://doi.org/10.1038/nphys2334}{{Experimental estimation of the dimension of classical and quantum systems}}, Nature Physics 8~(8) (2012) 588--591.
\newblock \href {https://doi.org/10.1038/nphys2334} {\path{doi:10.1038/nphys2334}}.
\newline\urlprefix\url{https://doi.org/10.1038/nphys2334}

\bibitem{AhrensNatPhys2012}
J.~Ahrens, P.~Badziag, A.~Cabello, M.~Bourennane, \href{http://dx.doi.org/10.1038/nphys2333}{{Experimental device-independent tests of classical and quantum dimensions}}, Nature Physics 8~(8) (2012) 592–595.
\newblock \href {https://doi.org/10.1038/nphys2333} {\path{doi:10.1038/nphys2333}}.
\newline\urlprefix\url{http://dx.doi.org/10.1038/nphys2333}

\bibitem{DivianszkySciRep2023}
P.~Divi{\'a}nszky, I.~M{\'a}rton, E.~Bene, T.~V{\'e}rtesi, \href{https://doi.org/10.1038/s41598-023-39529-0}{{Certification of qubits in the prepare-and-measure scenario with large input alphabet and connections with the Grothendieck constant}}, Scientific Reports 13~(1) (2023) 13200.
\newblock \href {https://doi.org/10.1038/s41598-023-39529-0} {\path{doi:10.1038/s41598-023-39529-0}}.
\newline\urlprefix\url{https://doi.org/10.1038/s41598-023-39529-0}

\bibitem{Grothendieck1953}
A.~Grothendieck, {R{\'e}sum{\'e} de la th{\'e}orie m{\'e}trique des produits tensoriels topologiques}, Bol.\ Soc.\ Mat.\ Sao Paulo 8 (1953) 1--79.

\bibitem{AcinPRA2006}
A.~Ac\'{\i}n, N.~Gisin, B.~Toner, \href{https://link.aps.org/doi/10.1103/PhysRevA.73.062105}{{Grothendieck's constant and local models for noisy entangled quantum states}}, Phys. Rev. A 73 (2006) 062105.
\newblock \href {https://doi.org/10.1103/PhysRevA.73.062105} {\path{doi:10.1103/PhysRevA.73.062105}}.
\newline\urlprefix\url{https://link.aps.org/doi/10.1103/PhysRevA.73.062105}

\bibitem{DivianszkyPRA2017}
P.~Divi\'anszky, E.~Bene, T.~V\'ertesi, \href{https://link.aps.org/doi/10.1103/PhysRevA.96.012113}{{Qutrit witness from the Grothendieck constant of order four}}, Phys. Rev. A 96 (2017) 012113.
\newblock \href {https://doi.org/10.1103/PhysRevA.96.012113} {\path{doi:10.1103/PhysRevA.96.012113}}.
\newline\urlprefix\url{https://link.aps.org/doi/10.1103/PhysRevA.96.012113}

\bibitem{BrunnerRevModPhys2014}
N.~Brunner, D.~Cavalcanti, S.~Pironio, V.~Scarani, S.~Wehner, \href{https://link.aps.org/doi/10.1103/RevModPhys.86.419}{{Bell nonlocality}}, Rev. Mod. Phys. 86 (2014) 419--478.
\newblock \href {https://doi.org/10.1103/RevModPhys.86.419} {\path{doi:10.1103/RevModPhys.86.419}}.
\newline\urlprefix\url{https://link.aps.org/doi/10.1103/RevModPhys.86.419}

\bibitem{PalazuelosJofMathPhys2016}
C.~Palazuelos, T.~Vidick, \href{https://doi.org/10.1063/1.4938052}{{Survey on nonlocal games and operator space theory}}, Journal of Mathematical Physics 57~(1) (2016) 015220.
\newblock \href {http://arxiv.org/abs/https://pubs.aip.org/aip/jmp/article-pdf/doi/10.1063/1.4938052/16729413/015220\_1\_online.pdf} {\path{arXiv:https://pubs.aip.org/aip/jmp/article-pdf/doi/10.1063/1.4938052/16729413/015220\_1\_online.pdf}}, \href {https://doi.org/10.1063/1.4938052} {\path{doi:10.1063/1.4938052}}.
\newline\urlprefix\url{https://doi.org/10.1063/1.4938052}

\bibitem{Raghavendra2009}
P.~Raghavendra, D.~Steurer, {Towards computing the Grothendieck constant}, in: Proceedings of the Twentieth Annual ACM-SIAM Symposium on Discrete Algorithms, SODA '09, Society for Industrial and Applied Mathematics, USA, 2009, p. 525–534.

\bibitem{FriezeCombinatorica1999}
A.~Frieze, R.~Kannan, \href{https://doi.org/10.1007/s004930050052}{{Quick Approximation to Matrices and Applications}}, Combinatorica 19~(2) (1999) 175--220.
\newblock \href {https://doi.org/10.1007/s004930050052} {\path{doi:10.1007/s004930050052}}.
\newline\urlprefix\url{https://doi.org/10.1007/s004930050052}

\bibitem{BorgsAdvinMath2008}
C.~Borgs, J.~Chayes, L.~Lovász, V.~Sós, K.~Vesztergombi, \href{https://www.sciencedirect.com/science/article/pii/S0001870808002053}{{Convergent sequences of dense graphs I: Subgraph frequencies, metric properties and testing}}, Advances in Mathematics 219~(6) (2008) 1801--1851.
\newblock \href {https://doi.org/https://doi.org/10.1016/j.aim.2008.07.008} {\path{doi:https://doi.org/10.1016/j.aim.2008.07.008}}.
\newline\urlprefix\url{https://www.sciencedirect.com/science/article/pii/S0001870808002053}

\bibitem{LinialCombinatorica2007}
N.~Linial, S.~Mendelson, G.~Schechtman, A.~Shraibman, \href{https://doi.org/10.1007/s00493-007-2160-5}{{Complexity measures of sign matrices}}, Combinatorica 27~(4) (2007) 439--463.
\newblock \href {https://doi.org/10.1007/s00493-007-2160-5} {\path{doi:10.1007/s00493-007-2160-5}}.
\newline\urlprefix\url{https://doi.org/10.1007/s00493-007-2160-5}

\bibitem{GohPRA2018}
K.~T. Goh, J.~m.~k. Kaniewski, E.~Wolfe, T.~V\'ertesi, X.~Wu, Y.~Cai, Y.-C. Liang, V.~Scarani, \href{https://link.aps.org/doi/10.1103/PhysRevA.97.022104}{Geometry of the set of quantum correlations}, Phys. Rev. A 97 (2018) 022104.
\newblock \href {https://doi.org/10.1103/PhysRevA.97.022104} {\path{doi:10.1103/PhysRevA.97.022104}}.
\newline\urlprefix\url{https://link.aps.org/doi/10.1103/PhysRevA.97.022104}

\bibitem{SliwaPhysLettA2003}
C.~Śliwa, \href{https://www.sciencedirect.com/science/article/pii/S0375960103011150}{Symmetries of the bell correlation inequalities}, Physics Letters A 317~(3) (2003) 165--168.
\newblock \href {https://doi.org/https://doi.org/10.1016/S0375-9601(03)01115-0} {\path{doi:https://doi.org/10.1016/S0375-9601(03)01115-0}}.
\newline\urlprefix\url{https://www.sciencedirect.com/science/article/pii/S0375960103011150}

\bibitem{AraujoArxiv2024}
M.~Araújo, I.~Klep, A.~J.~P. Garner, T.~Vértesi, M.~Navascués, \href{https://arxiv.org/abs/2311.18707}{First-order optimality conditions for non-commutative optimization problems} (2024).
\newblock \href {http://arxiv.org/abs/2311.18707} {\path{arXiv:2311.18707}}.
\newline\urlprefix\url{https://arxiv.org/abs/2311.18707}

\bibitem{Pacheco2011}
P.~Pacheco, \href{https://books.google.hu/books?id=SEmfraJjvfwC}{{An Introduction to Parallel Programming}}, Elsevier Science, 2011.
\newline\urlprefix\url{https://books.google.hu/books?id=SEmfraJjvfwC}

\bibitem{HanCUDA2019}
J.~Han, B.~Sharma, \href{https://books.google.hu/books?id=dhWzDwAAQBAJ}{{Learn CUDA Programming: A beginner's guide to GPU programming and parallel computing with CUDA 10.x and C/C++}}, Packt Publishing, 2019.
\newline\urlprefix\url{https://books.google.hu/books?id=dhWzDwAAQBAJ}

\bibitem{BitnerCACM1976}
J.~R. Bitner, G.~Ehrlich, E.~M. Reingold, {Efficient generation of the binary reflected Gray code and its applications}, Communications of the ACM 19~(9) (1976) 517--521.

\bibitem{ConderDiscMath1999}
M.~Conder, \href{https://www.sciencedirect.com/science/article/pii/S0012365X98001885}{{Explicit definition of the binary reflected Gray codes}}, Discrete Mathematics 195~(1) (1999) 245--249.
\newblock \href {https://doi.org/https://doi.org/10.1016/S0012-365X(98)00188-5} {\path{doi:https://doi.org/10.1016/S0012-365X(98)00188-5}}.
\newline\urlprefix\url{https://www.sciencedirect.com/science/article/pii/S0012365X98001885}

\bibitem{SankarPSISPCS2004}
K.~Sankar, V.~Pandharipande, P.~Moharir, {Generalized Gray codes}, in: Proceedings of 2004 International Symposium on Intelligent Signal Processing and Communication Systems, 2004. ISPACS 2004., 2004, pp. 654--659.
\newblock \href {https://doi.org/10.1109/ISPACS.2004.1439140} {\path{doi:10.1109/ISPACS.2004.1439140}}.

\bibitem{HerterTCS2018}
F.~Herter, G.~Rote, \href{https://www.sciencedirect.com/science/article/pii/S0304397517308472}{{Loopless Gray code enumeration and the Tower of Bucharest}}, Theoretical Computer Science 748 (2018) 40--54, fUN with Algorithms.
\newblock \href {https://doi.org/https://doi.org/10.1016/j.tcs.2017.11.017} {\path{doi:https://doi.org/10.1016/j.tcs.2017.11.017}}.
\newline\urlprefix\url{https://www.sciencedirect.com/science/article/pii/S0304397517308472}

\bibitem{GrucaPRA2010}
J.~Gruca, W.~Laskowski, M.~\ifmmode~\dot{Z}\else \.{Z}\fi{}ukowski, N.~Kiesel, W.~Wieczorek, C.~Schmid, H.~Weinfurter, \href{https://link.aps.org/doi/10.1103/PhysRevA.82.012118}{Nonclassicality thresholds for multiqubit states: Numerical analysis}, Phys. Rev. A 82 (2010) 012118.
\newblock \href {https://doi.org/10.1103/PhysRevA.82.012118} {\path{doi:10.1103/PhysRevA.82.012118}}.
\newline\urlprefix\url{https://link.aps.org/doi/10.1103/PhysRevA.82.012118}

\bibitem{GondzioJCompApplMath2014}
J.~Gondzio, J.~A. Gruca, J.~J. Hall, W.~Laskowski, M.~Żukowski, \href{https://www.sciencedirect.com/science/article/pii/S0377042713006730}{Solving large-scale optimization problems related to bell’s theorem}, Journal of Computational and Applied Mathematics 263 (2014) 392--404.
\newblock \href {https://doi.org/https://doi.org/10.1016/j.cam.2013.12.003} {\path{doi:https://doi.org/10.1016/j.cam.2013.12.003}}.
\newline\urlprefix\url{https://www.sciencedirect.com/science/article/pii/S0377042713006730}

\bibitem{Arajo2020}
M.~Araújo, F.~Hirsch, M.~T. Quintino, \href{http://dx.doi.org/10.22331/q-2020-10-28-353}{Bell nonlocality with a single shot}, Quantum 4 (2020) 353.
\newblock \href {https://doi.org/10.22331/q-2020-10-28-353} {\path{doi:10.22331/q-2020-10-28-353}}.
\newline\urlprefix\url{http://dx.doi.org/10.22331/q-2020-10-28-353}

\bibitem{MontinaEntropy2019}
A.~Montina, S.~Wolf, \href{http://dx.doi.org/10.3390/e21020104}{Discrimination of non-local correlations}, Entropy 21~(2) (2019) 104.
\newblock \href {https://doi.org/10.3390/e21020104} {\path{doi:10.3390/e21020104}}.
\newline\urlprefix\url{http://dx.doi.org/10.3390/e21020104}

\bibitem{Designolle2023}
S.~Designolle, G.~Iommazzo, M.~Besan\ifmmode~\mbox{\c{c}}\else \c{c}\fi{}on, S.~Knebel, P.~Gel\ss{}, S.~Pokutta, \href{https://link.aps.org/doi/10.1103/PhysRevResearch.5.043059}{Improved local models and new bell inequalities via frank-wolfe algorithms}, Phys. Rev. Res. 5 (2023) 043059.
\newblock \href {https://doi.org/10.1103/PhysRevResearch.5.043059} {\path{doi:10.1103/PhysRevResearch.5.043059}}.
\newline\urlprefix\url{https://link.aps.org/doi/10.1103/PhysRevResearch.5.043059}

\bibitem{Designolle2024}
S.~Designolle, T.~V\'ertesi, S.~Pokutta, \href{https://link.aps.org/doi/10.1103/PhysRevA.109.022205}{Symmetric multipartite bell inequalities via frank-wolfe algorithms}, Phys. Rev. A 109 (2024) 022205.
\newblock \href {https://doi.org/10.1103/PhysRevA.109.022205} {\path{doi:10.1103/PhysRevA.109.022205}}.
\newline\urlprefix\url{https://link.aps.org/doi/10.1103/PhysRevA.109.022205}

\end{thebibliography}

\end{document}